\documentclass[letterpaper,conference]{IEEEtran}
\usepackage{graphicx}
\usepackage{amsmath}
\usepackage{amssymb}

\begin{document}

\title{Compressed Sensing based Protocol for Efficient Reconstruction of Sparse Superimposed Data in a Multi-Hop Wireless Sensor Network}
\author{\authorblockN{Megumi Kaneko$^{\*}$ and Khaldoun Al Agha$^{\#}$}\\
\authorblockA{$^{\*}$Graduate School of Informatics, Kyoto University,\\
Yoshida Honmachi Sakyo--ku, Kyoto, 606-8501, Japan\\
Email: meg@i.kyoto-u.ac.jp}
$^{\#}$LRI, Universit\'{e} Paris-Sud, 91405 Orsay cedex, France\\
Email: alagha@lri.fr}

\maketitle

\begin{abstract}
We consider a multi-hop wireless sensor network that measures sparse events and propose a simple forwarding protocol based on Compressed Sensing (CS) which does not need any sophisticated Media Access Control (MAC) scheduling, neither a routing protocol, thereby making significant overhead and energy savings. By means of flooding, multiple packets with different superimposed measurements are received simultaneously at any node. Thanks to our protocol, each node is able to recover each measurement and forward it while avoiding cycles. Numerical results show that our protocol achieves close to zero reconstruction errors at the sink, while greatly reducing overhead. This initial research reveals a new and promising approach to protocol design through CS for wireless mesh and sensor networks.

\end{abstract}

\section{Introduction}

Wireless sensor networks are used in large sensing areas where multi-hop routing is needed to forward the information from the sensors to the sinks. When a sensor wakes up according to an event and then measures a physical value, it first needs to schedule its transmission in an interference-free slot using a Media Access Control (MAC) layer scheme and then to use a specific routing protocol in order to fix the next hop before reaching the final sink. Those steps include heavy operations for small sensors and consume a big quantity of energy. However, they are mandatory in order to avoid collisions and to determine the best routes to reach sinks.


Recently, the groundbreaking theory of Compressed Sensing (CS) was developed, which stipulates that a vector with correlated entries, i.e., that can be transformed into a sparse vector through a transformation basis, can be recovered with high accuracy from a few random projections onto another, incoherent basis~\cite{Can06feb}~\cite{Don06apr}~\cite{Can06dec}. CS, which has been widely used in the domains of digital signal and image processing, is envisioned as a highly promising tool for improving the performance of wireless communication networks.
In particular, based on the decentralized compression techniques presented in~\cite{Hau08mar}, there has been a number of works exploiting CS in wireless sensor networks dealing with space-time correlated data as in~\cite{Que09feb}\cite{Wan10oct}. In these works, the underlying assumption is that all sensor nodes have a data to send which are gathered at the sink and reconstructed with few observations using CS techniques. The transmission of each data follows some predetermined routes, such that packets are received without errors. In~\cite{Ngu10oct}, the CS technique is combined to the well-known Network Coding (NC), which provides an efficient method for data communication in wireless networks, as multiple packets aggregated in a single packet are decoded using prior information. Introducing CS enabled to mitigate the problem of exploding header size with NC where the header grew proportionally to the number of aggregated packets, since it became possible to decode the salient information from fewer number of packets. Note also that, as a consequence, CS enables to significantly decrease the delay required by NC for gathering packets, as at least $n$ different combined packets were needed to decode $n$ packets. However, in this work too, the transmissions follow predetermined routes in order to avoid any collision, i.e., packets are aggregated from node to node and not
``over-the-air".
By contrast,~\cite{Men09mar} considers a one-hop sensor network where multiple nodes may transmit their measurement simultaneously to the sinks. As the measured events are assumed to be sparse, i.e., the number of measurements simultaneously received at the sink is much smaller than the total number of sensors, the sink is able to recover each measurement from the superimposed signals from few observations, thanks to CS algorithms.
Such an approach is also taken for developing new multiple access schemes for random data traffic in~\cite{Mao10apr}, for channel reservation in~\cite{Qas09sep}, and for downlink scheduling in~\cite{Bha09jun}.

In this work, we consider a multi-hop wireless sensor network and take the approach of~\cite{Men09mar}--\cite{Bha09jun} where the events to be reported occur sparsely. Each sensor with a value to report combines it with its signature sequence giving its Identity (ID) and broadcasts the resulting vector to its neighbor nodes. We develop a protocol based on CS and flooding that enables the sink to obtain and reconstruct the sparse measurement data with high accuracy without any heavy routing nor MAC protocol. In a one-hop network, the simultaneously received measurements may be resolved by the CS-based algorithms in~\cite{Men09mar}--\cite{Bha09jun}, but they pose major problems in a multi-hop network where every node forwards all received packets, as the number of interfering measurement packets may be increased drastically, thereby causing the CS-based algorithms to perform poorly due to loss of sparsity. Moreover, if different measurements of a same source but generated at different times are contained in the same packet, they would be hardly resolvable.
To alleviate the aforementioned issues, in our protocol, the length of the signature sequences is further increased so that each sequence can identify the source node and time stamp, which allows to avoid cycles in the flooding process as follows. When neighboring nodes receive the superimposed packets, they first recover each measurement data from an under-determined system of equations, using a CS algorithm based on $\ell_1$-$\ell_2$ optimization, commonly used in the case of noisy measurements~\cite{Elad10}. After decoding each data, they check in their local tables if it was already sent, using the recovered origin node ID and time stamp from its sequence; if it is the case, they discard the data. If not, they again superimpose each data into a unique packet which is broadcasted to their neighbors. This process is repeated until reaching the sink. In addition to avoiding cycles, this process allows to maintain the "sparsity" of the superimposed data by discarding superfluous packets, thereby guaranteeing a good performance of the CS algorithms. Still, the length of these sequences is far below the total number of possible combinations of IDs and time stamps, which would be required in a CDMA-based protocol.

Many advantages are offered by our new protocol:

\begin{itemize}
\item Sensors do not need any MAC scheduling that consumes resources, energy and increase the delivery delays.
\item Sensors do not need any routing protocols that need many control messages that waste the bandwidth and also increase the energy consumption.
\item Decoding in the sensors needs processing power and consumes energy. However, those operations represent a negligible consumption compared to the radio reception, listening, transmission used by the MAC scheduling.
\item The time stamp represents an additional overhead but is bounded by, e.g., the maximum number of hops in a network. We show that it represents a negligible part compared to the overhead induced by a routing protocol and even compared to a flooding-based protocol with CDMA.
\item While discarding packets that have already been forwarded for keeping sparsity of the superimposed data and avoiding cycles, our protocol still offers diversity for each measurement data as a packet may reach the sink by many routes. Even if a packet is lost, the sink or the intermediate nodes may receive it by other routes.
\end{itemize}

The paper is organized as follows: after giving some basic results in Compressed Sensing and presenting related works, we explain the system model and our proposed protocol. The simulation results show its effectiveness, while reducing the amount of overhead compared to conventional protocols.

\section{Results from Compressed Sensing}
\label{sec CS}

Compressed Sensing is a newly developed mathematical theory that enables to solve underdetermined systems of equations under the sparsity prior of the solution, a problem formulated as
\begin{equation}\label{eq P0}
(P_{0}): \quad \min_{\mathbf{x}} \|\mathbf{x}\|_0, \quad \text{subject to} \quad \mathbf{y}=\mathbf{A}\mathbf{x},
\end{equation}
where $\mathbf{x} \in \mathbb{R}^N$, $\mathbf{y} \in \mathbb{R}^M$, $\mathbf{A} \in \mathbb{R}^{M\times N}$ with $M<N$, and $\|\mathbf{x}\|_0=\text{card}\{i, x_i \neq 0\}$, where sparsity of $\mathbf{x}$ is defined by $\|\mathbf{x}\|_0=K << N$.
However, this is a non-convex optimization problem that requires combinatorial search over all possible sparse combinations of columns of $\mathbf{A}$, which rapidly becomes intractable. Therefore, convex relaxation of this problem has been considered by the following $\ell_1$-minimization,
\begin{equation}\label{eq P1}
(P_{1}): \quad \min_{\mathbf{x}} \|\mathbf{x}\|_1, \quad \text{subject to} \quad \mathbf{y}=\mathbf{A}\mathbf{x},
\end{equation}
where $\|\mathbf{x}\|_1=\sum_{i=1}^n|x_i|$, for which various efficient algorithms have been developed, among which Pursuit algorithms and Iterative Shrinkage Algorithms~\cite{Elad10}.

Under the assumption of noisy observations, the optimization problem may be reformulated as
\begin{equation}\label{eq P1e}
(P_{1}^{\epsilon}): \quad \min_{\mathbf{x}} \|\mathbf{x}\|_1, \quad \text{subject to} \quad \|\mathbf{y}-\mathbf{A}\mathbf{x}\|_2 \leq \epsilon,
\end{equation}
which can be equivalently expressed as an $\ell_1-\ell_2$ minimization problem,
\begin{equation}\label{eq P12}
(P_{1-2}^{\lambda}): \quad \min_{\mathbf{x}} \lambda\|\mathbf{x}\|_1 + \frac{1}{2}\|\mathbf{y}-\mathbf{A}\mathbf{x}\|_2^2,
\end{equation}
where the parameter $\lambda$ can be viewed as a Lagrange Multiplier making a trade-off between representation error and sparsity of the solution.
It was shown that random matrices $\mathbf{A}$ whose entries are drawn from Gaussian or Sub-Gaussian (e.g., Bernouilli) distributions, guarantee stable recovery in the noisy case, provided that the number of measurements obey
\begin{equation}
    M \geq cK\log(N/K),
\end{equation}
where $c$ is a constant (this condition guarantees that $\mathbf{A}$ will guarantee the Restricted Isometry Property (RIP) with high probability, enabling stable recovery by $\ell_1$-minimization, see~\cite{Elad10} and references therein).
In our proposed protocol, we will employ the Iterative Shrinkage-Thresholding Algorithm (ISTA)~\cite{Dau04nov} as it efficiently solves the
$\ell_1-\ell_2$ minimization $(P_{1-2}^{\lambda})$ with very low complexity.


\section{Related work}
\label{sec RelatedWork}

In the multiple access schemes of~\cite{Qas09sep}\cite{Mao10apr}\cite{Bha09jun} for a single-hop system, several Mobile Stations (MS) access the Base Station (BS) at the same time, on the same frequency.
Assuming $J$ MSs in total, each MS $j$ is assigned a pseudo-random signature sequence $\mathbf{a}_j$, a vector of size $M$, which entries are generated by Bernoulli random variables of probability 1/2, with values $\pm 1$. Thus, in the absence of noise, the received signal $\mathbf{y}$ at the BS may be expressed as
\[
\mathbf{y}=
  \begin{bmatrix}
    \mathbf{a}_1 & \hdots & \mathbf{a}_j & \hdots  & \mathbf{a}_J\\
  \end{bmatrix}\mathbf{v}
  =\mathbf{A}\mathbf{v},
 \]
where $\mathbf{v}$ is the vector of size $J$ with component $\{\mathbf{v}\}_j=1$ if MS $j$ transmitted, and $\{\mathbf{v}\}_j=0$ if not. As the number of transmitting MSs is assumed to be much smaller than the total number of MSs $J$, $\mathbf{v}$ is a sparse vector. Hence, $\mathbf{v}$ is recovered at the BS using CS algorithms solving problems (\ref{eq P1}), or (\ref{eq P12}) in the noisy case.
Note that, the usage of these pseudo-random signature sequences to identify each MS may be regarded as similar to a Code Division Multiple Access (CDMA) system. However, CDMA systems typically require orthogonal codes to differentiate each MS, which would require each sequence $\mathbf{a}_j$ to be at least of size $J$, whereas these multi-access systems with CS deal with non-orthogonal sequences since $M<<J$, which can be viewed as an overloaded CDMA system, in which case traditional CDMA receivers such as the linear minimum mean squared error filter perform very badly, as shown in~\cite{Bha09jun}.
Nevertheless, we will also compare our proposed protocol to a CDMA-based reference protocol that does not require any routing.

\section{System Model}
\label{sec SysMod}

\begin{figure}[h]
\begin{center}
   \includegraphics[width=.55\linewidth]{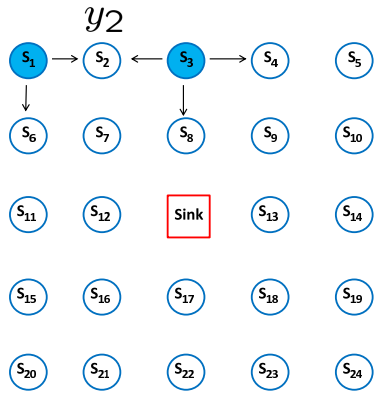}
\caption{Multi-Hop Wireless Sensor Network Model, Step 1}
\label{fig:lattice1}
\end{center}
\end{figure}

We consider a multi-hop wireless sensor network with $N$ nodes and one sink, forming a lattice as shown in Fig. \ref{fig:lattice1}. Each sensor node $S_n$ can be either in transmit or receive mode at any one time. If a sensor detects an event, the network forwards the corresponding measurement to the sink based on the proposed protocol. As in~\cite{Men09mar} which considers a one-hop sensor network (but multiple sinks), we assume that there may be up to $K$ events occurring simultaneously within the whole network, but these events are considered to be sparse compared to the number of nodes, namely $K << N$.


We assume digital transmissions as in~\cite{Men09mar}, i.e., measurements $x \in \{-1,1\}$, instead of directly transmitting scalar values by amplitude modulation as in~\cite{Hau08mar}, since interfering analog signals pose difficult problems in a multi-hop network. For example, a node could simultaneously receive two packets from different paths, but containing the same measurement $x$, in which case it would erroneously decode $2x$. As the goal here is to present the benefits of our protocol, we consider the digital case for simplicity, but the protocol will be developed to accommodate analog transmissions in the next phase.

Given the lattice structure of the network, it is assumed that each node only communicates with its closest neighbors at distance $d$, so there may be $2$, $3$ or $4$ neighbors depending on the node's location. For example, a packet sent by node $S_1$ in Fig. \ref{fig:lattice1} will be received by nodes $S_2$ and $S_6$ only, while a packet sent by node $S_7$ will be received by nodes $S_2$, $S_6$, $S_8$ and $S_{12}$.
For simplicity, all channels are considered to be Additive White Gaussian Noise (AWGN), and all links between neighbors at distance $d$ have the same gain. Channel fading and path loss effects will be considered in the future work\footnote{Note that, although fading and path loss effects are considered in~\cite{Men09mar}, the channel gains between each sensor and sink are assumed to be known perfectly as they form the columns of sensing matrix $\mathbf{A}$.}. Note that all nodes are synchronized as in~\cite{Hau08mar}\cite{Men09mar}, but there are interesting design issues in the asynchronous case, to which the protocol may be extended.


\begin{figure}[h]
\begin{center}
   \includegraphics[width=.55\linewidth]{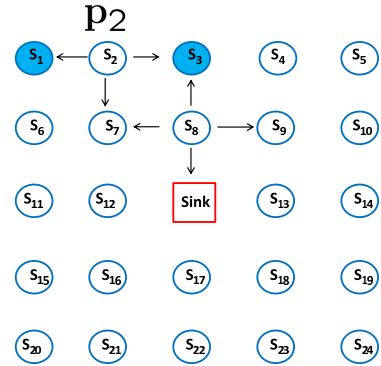}
\caption{Multi-Hop Wireless Sensor Network Model, Step 2}
\label{fig:lattice}
\end{center}
\end{figure}

\section{Proposed Protocol}
\label{sec Algo}

The proposed routing protocol consists in a simple flooding whereby upon packet reception, each node broadcasts this packet locally and then all its neighbors forward this information to their neighbors and so on, until it reaches the sink. In this setting, each packet will be forwarded through different paths, creating multiple copies that provide diversity at the sink. As in the works presented in Section \ref{sec RelatedWork}, we use in our system the Bernoulli sequences for identification, forming the independent projection vectors required for recovering the compressed signals.
However, if each source node is only assigned a unique signature sequence as in the previous works, it will not be possible to distinguish whether the multiple received packets are due to self-interference, i.e., they are copies of a same original packet, or whether they correspond to measurements from the same source but taken at different times.

To alleviate this problem, we associate each measurement with a unique sequence $\mbox{\bf a}_{n,l}$, vector of size $M$, that identifies the pair (origin node ID $S_n$, number of hops $l$), where $S_n$ is the sensor that originally detected this measurement, and $l$ is its time stamp parameter, defined as the number of hops elapsed by forwarding the packet bearing this measurement, counted from its origin node up to its current location in the network.
We have $\mbox{\bf a}_{n,l} \in \{-1,+1\}^M$, given by a pseudo-random noise signature sequence generated by Bernoulli random variables of probability 1/2, with $\pm 1$ values. Thus, fixing a maximum delay for a packet or time to live $L$ in number of hops, which depends on the network size, there are $N L$ sequences $\mbox{\bf a}_{n,l}$ of size $M$, where we set $M< NL$ for reducing overhead consumption. These sequences are assumed to be known at each node.
They are gathered into matrix $\mathbf{A}$ of size $M \times NL$, whose column $q=(n-1)L+l$ contains $\mbox{\bf a}_{n,l}$.
For notational convenience, let $\mathbf{x}$ denote the $NL$-length sparse vector where $\{\mathbf{x}\}_{(n-1)L+l}=x_{k,l}$ which groups all the measurements with possible hop counts, i.e., $\mathbf{x}=[x_{1,0}...x_{1,L}...x_{n,l}...x_{N,0}...x_{N,L}]^T$ where the $L+1$ first terms store the measurements of source node $S_1$ for $l=0..L$ and similarly, the $L+1$ last terms for source node $S_N$ for $l=0..L$.

Then, the proposed algorithm works as follows:

\noindent \textbf{1.} If source node $S_j$ in transmit mode detects a new event with measurement value $x_{j,0} \in \{-1,+1\}$ initialized by $l=0$, he transmits the packet of $M$ bits
\begin{equation}
    p_{j}=x_{j,0}\mbox{\bf a}_{j,0},
\end{equation}
which is received by all its neighbor nodes $k$ $\in$ $\mbox{\bf N}(S_j)$.
For example, assuming $S_1$ and $S_3$ to be source nodes in Fig. \ref{fig:lattice1}, packets $p_{1}=x_{1,0}\mbox{\bf a}_{1,0}$ and $p_{3}=x_{3,0}\mbox{\bf a}_{3,0}$ are received by their respective neighbors, i.e. $S_2, S_6$ for $S_1$ and $S_2, S_4, S_{8}$ for $S_3$.

\noindent \textbf{2.} At a time $t$, if sensor node $S_k$ is in receive mode, he receives signal $\mathbf{y}_{k}$ containing all simultaneous transmissions from his neighbor nodes $n$ $\in$ $\mbox{\bf N}(S_k)$. Thus, the received signal at a node $S_k$ is written
  \begin{equation}\label{eq 1}
     \mathbf{y}_{k}=\sum_{n \in \mathbf{N}(S_k)}p_{n}=\sum_{j,l}x_{j,l}\mathbf{a}_{j,l}+\mathbf{z}_{k},
  \end{equation}
where $x_{j,l}$, the measurement with source node $S_j$ is included in $p_{n}$ if it was actually forwarded by node $S_n$. Vector $\mathbf{z}_k$ of size $M$, denotes the AWGN.
Using the matrix $\mathbf{A}$ defined above, this is equivalently reformulated as
\begin{equation}
\mathbf{y}_k=\mathbf{A}\mathbf{x}_k+\mathbf{z}_k,
\end{equation}
where $\mathbf{x}_k$ is the vector of size $NL$ defined as
\[
   \{\mathbf{x}_k\}_i = \left\{
  \begin{array}{l l}
     \{\mathbf{x}\}_i & \quad \text{if $\{\mathbf{x}\}_i=x_{j,l}$ is received by $S_k$}\\
     0 & \quad \text{otherwise}\\
   \end{array} \right.
 \]
In the example of Fig. \ref{fig:lattice1}, note that $S_j=S_n$, i.e., the source nodes and transmit nodes coincide since we describe Step 1, so the received signal at receiver node $S_2$ is written $\mathbf{y}_2=\mathbf{A}\mathbf{x}_2+\mathbf{z}_2$, where
$\mathbf{x}_2=[x_{1,0} \ 0...0 \ x_{3,0} \ 0...0]^T$.



\noindent \textbf{3.} Receiver node $S_k$ decodes each superimposed measurements contained in the received signal $\mathbf{y}_{k}$, using CS principles. In particular, this problem can be formulated by $\ell_1$-$\ell_2$ optimization as in (\ref{eq P12}),
     \begin{equation}\label{eq 4}
  \mathbf{\tilde{x}}_k=\arg \min_{\mathbf{x}_k} \frac{1}{2}\|\mathbf{y}_{k}-\mathbf{A}\mathbf{x}_k\|_2^2+\lambda \|\mathbf{x}_k\|_1,
  \end{equation}
which we solve by ISTA with very good reconstruction abilities with low complexity. 

\noindent \textbf{4.} The reconstructed vector $\mathbf{\tilde{x}}_k$ is renormalized so that each component belongs to $\{-1,+1\}$.

\noindent \textbf{5.} After decoding the received measurements at node $S_k$, their respective (source ID, time stamp)-sequence $\mbox{\bf a}_{j,l}$ could be identified.
Node $S_k$ then compares the sequence $\mbox{\bf a}_{j,l}$ of each newly decoded measurement to the sequences of previously received ones that are stocked in its local table $\mathbf{Q}(S_k)=\{\mbox{\bf a}_{q,l'} \ \text{received by} \ S_k\}$.
      \begin{itemize}
        \item if there exists a sequence $\mbox{\bf a}_{q,l'}$ in $\mathbf{Q}(S_k)$ with $q=j$ and $l' \leq l$: the same data was already received before with high probability, since no fading is assumed, such that any new measurement should have a hop count $l < l'$, unless the packet that followed the shortest path was lost. However, this will rarely occur in this setting. Thus, $S_k$ discards the decoded measurement $\tilde{x}_{j,l}\mbox{\bf a}_{j,l}$, from the set of data to be forwarded next.
        \item if there exists a sequence $\mbox{\bf a}_{q,l'}$ in $\mathbf{Q}(S_k)$ with $q=j$ and $l'>l$: although both share the same origin node $S_j$, this is a new data as it has a smaller hop count. Therefore, $S_k$ will forward the decoded data $\mathbf{\tilde{x}}_{j,l}\mbox{\bf a}_{j,l}$.
        \item in all other cases, $S_k$ will forward $\tilde{x}_{j,l}\mbox{\bf a}_{j,l}$.
      \end{itemize}

\noindent \textbf{6.} For all data $\tilde{x}_{j,l}\mbox{\bf a}_{j,l}$ to be forwarded, the hop count is incremented to $l+1$.
All decoded measurements as well as sequences in $\mathbf{Q}(S_k)$ for which $l>L$ are discarded.
Then, $S_k$ superimposes all the remaining data to be forwarded into a packet $p_{k}$ composed of $M$ bits,
\begin{equation}
    \mathbf{p}_k=\mathbf{A}\mathbf{\tilde{x}}_k=\sum_{j,l}\tilde{x}_{j,l}\mathbf{a}_{j,l},
\end{equation}
and forwards it. Thus, in Fig. \ref{fig:lattice}, node $S_2$ sends $p_2=\tilde{x}_{1,1}\mbox{\bf a}_{1,1}+\tilde{x}_{3,1}\mbox{\bf a}_{3,1}$.

\noindent \textbf{7.} The sink node runs the reconstruction algorithm for all incoming packets during the session of duration $T_{out}$. Whenever multiple versions of the same data are received, more diversity gain may be achieved to improve the decision accuracy.

\section{Numerical Results}

We first evaluate our proposed protocol in a network of $25$ nodes. In a session, $K$ sources generate measurements that are forwarded to the sink as explained in Section \ref{sec Algo}. These sparse generation events occur at random times during each session, e.g., they may occur simultaneously. Each session ends after time out $T_{\mathrm{out}}$ in hop counts, at which point the measurements collected by the sink are evaluated. The Signal-to-Noise Ratio (SNR) between neighboring nodes is fixed to $10$ dB. Each simulated point is averaged over $300$ sessions.
We evaluate the reconstruction error as in~\cite{Que09feb}\cite{Wan10oct}, averaged over all sessions, defined for one session as
\begin{equation}\label{eq NMSE}
    \epsilon=\frac{\|\bf \hat{x}-\mbox{\bf x}_0\|_2}{\|\mbox{\bf x}_0\|_2},
\end{equation}
where $\mbox{\bf x}_0$ is the vector of size $N$ containing the original measurements for all $N$ sensors, i.e., its corresponding components are in $\{-1,+1\}$ if sensor $S_j$ is an origin node, and zero otherwise, while $\bf \hat{x}$ denotes the reshaped vector of collected measurements at the sink at the end of a session.

\begin{figure}[h]
\begin{center}
   \includegraphics[width=1\linewidth]{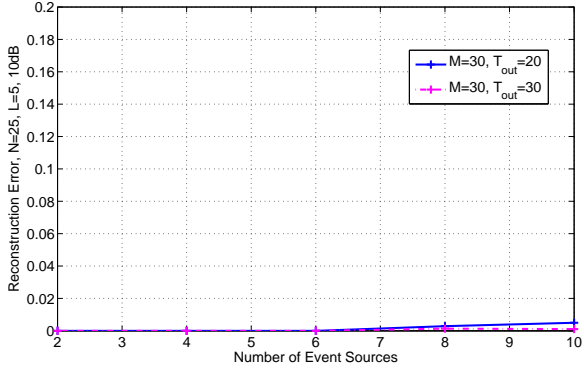} 
\vspace{-0.6cm}
\caption{Proposed Protocol: Reconstruction Error performance for different values of $T_{out}$, $M=30$, $N=25$, $L=5$.}
\label{fig:REN25}
\end{center}
\end{figure}

\begin{figure}[h]
\begin{center}
   \includegraphics[width=1\linewidth]{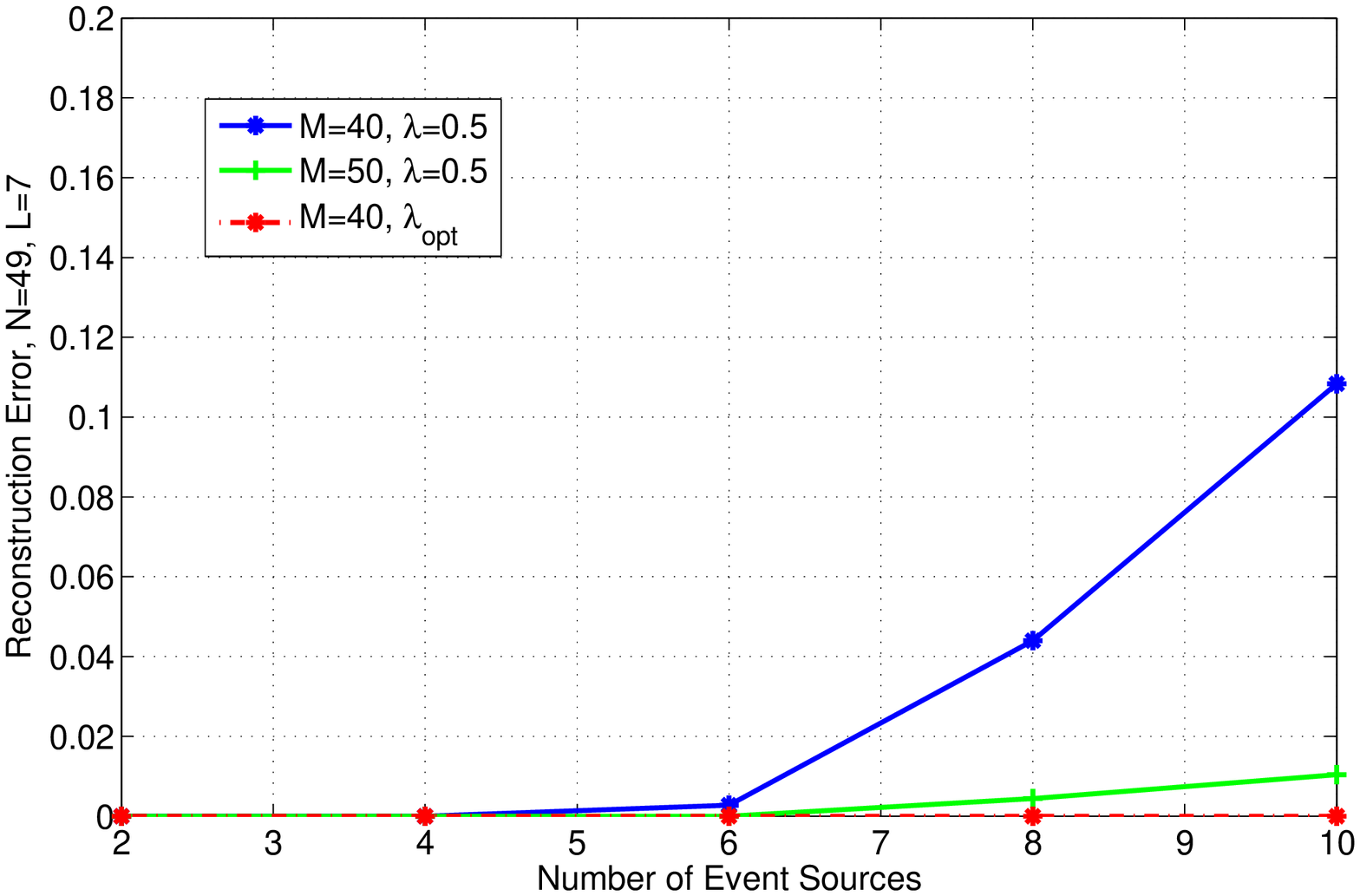} 
\vspace{-0.6cm}
\caption{Proposed Protocol: Reconstruction Error performance for different values of $M$, $\lambda$ and $T_{out}=30$, $N=49$, $L=7$.}
\label{fig:REN49}
\end{center}
\end{figure}

\begin{figure}[h]
\begin{center}
   \includegraphics[width=1\linewidth]{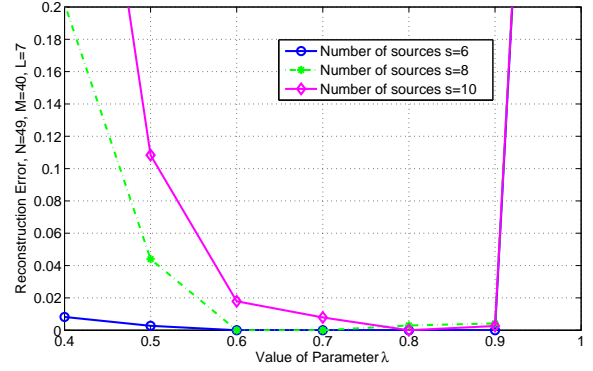} 
\vspace{-0.6cm}
\caption{Proposed Protocol: Reconstruction Error performance for different values of $\lambda$ and $s=6, 8, 10$; $T_{out}=30$, $N=49$, $M=40$ and $L=7$.}
\label{fig:REN49Lambda}
\end{center}
\end{figure}

Fig. \ref{fig:REN25} shows the average reconstruction error of our protocol for $K =2,...,10$, for sequence length $M=30$, $L=5$ and $T_{\mathrm{out}}=20,30$. We fix $\lambda=0.5$ in the decoding algorithm by ISTA. We observe that in both cases, excellent reconstruction performance is achieved for all numbers of event sources $K$, even for larger values of $K$ which imply lower sparsity in the number of superimposed measurement data. The error performance is improved by increasing $T_{\mathrm{out}}$, as more diversity gain is achieved at the expense of delay.

Next, we consider a network of $N=49$ nodes for $T_{\mathrm{out}}=30$, $L=7$ and $M=40, 50$.
For fixed $\lambda=0.5$, the reconstruction error increases as $K$ increases, due to lower sparsity: as the number of interfered packets increases, it becomes more difficult to correctly receive and recover all measurements at each node. However, by increasing $M$ from $40$ to $50$, the error performance decreases significantly for $K=8, 10$, since it provides a larger number of independent observations that allow to correctly decode a higher number of superimposed data by improving the estimation accuracy of the CS algorithm. Note that, even for small values of $K$, the number of interfering packets to be decoded at a certain node may be much more, since multiple copies of a given packet may be received through different paths, reducing the sparsity of the received signal $\mathbf{y}_{k}$.
However, increasing $M$ creates a larger overhead, even though it is far smaller than what is required by conventional protocols, as shown later.
Instead of increasing $M$, Fig. \ref{fig:REN49} shows that the reconstruction error can be made close to zero by adapting the parameter $\lambda$ to $\lambda_{opt}$, given the number of event sources, namely the sparsity of the system. Thus,
Fig. \ref{fig:REN49Lambda} evaluates the reconstruction error performance for the number of event sources $s=6, 8, 10$, and varying values of $\lambda$. Note that curves for the "sparser" cases $s=2, 4$ are not represented, as their reconstruction error stays close to zero for $\lambda \in [0.4,...,0.9]$. We observe that, as the number of sources increases, and hence as the sparsity diminishes, the reconstruction performance by ISTA becomes more sensitive to an accurate optimization of $\lambda$. In this case, the optimized values of $\lambda$ could be chosen as $\lambda_{opt}=0.6, 0.7, 0.8$ for $s=6, 8, 10$, respectively. The tendency for $\lambda_{opt}$ to increase with $s$ can be interpreted by the fact that more weight should be put on the $\ell_1$ term in (\ref{eq P12}) for finding a sparse solution, as more and more "denser" candidate solutions appear.
Of course, the excellent reconstruction performance achieved by $\lambda_{opt}$ comes at the price of increased computational complexity at each node, even though it should remain at a reasonable level given the very low computational complexity of ISTA. On the other hand, the alternative to optimizing $\lambda$ is to sufficiently increase $M$, which also provides very good performance as already shown in Fig. \ref{fig:REN49} with $M=50$, $\lambda=0.5$. The trade-off is thus between algorithm complexity or increased overhead, which is evaluated next.

\begin{figure}[h]
\begin{center}
   \includegraphics[width=1\linewidth]{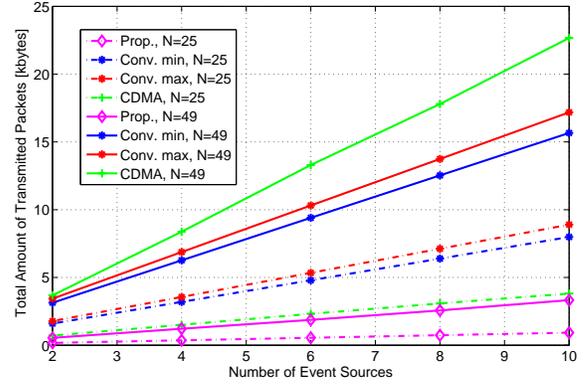} 
\caption{Average amount of transmitted packets per session in [kbytes] for two cases: $N=25$, $M=30$, $L=5$, $T_{out}=30$ and $N=49$, $M=50$, $L=7$, $T_{out}=30$.}
\label{fig:Bits}
\end{center}
\vspace{-0.6cm}
\end{figure}

We compare the proposed protocol with two reference algorithms: a Conventional (Conv.) algorithm based on AODV routing~\cite{aodv} and MAC scheduling, and a flooding-based protocol similar to our proposed algorithm, but using CDMA as explained in Section \ref{sec RelatedWork}. The comparison is only made in terms of overhead, since both reference algorithms should ideally provide excellent reconstruction performances under this system model, given the absence of interference for the first one, and due to the orthogonal codes for the second.
As overhead and data (origin node ID and measurement) are not separable in the proposed protocol, we evaluate the total amount of packets generated by each method. For the Conv. algorithm, the bounds are roughly given by
\begin{equation}
B_{AODV}=B_{RREQ}(N-1) + B_{RREP} h,
\end{equation}
where $B_{RREQ}=32$ bytes for Route Request, $B_{RREP}=28$ bytes for Route Reply, and
 \begin{equation}
B_{MAC}=(b_{ID+DATA}+b_{CRC}) h=2.5h,
\end{equation}
where $b_{ID+DATA}=11$ bits and $b_{CRC}=8$ bits for Cyclic Redundancy Check. The lower bound $B_{Conv}^{min}$ is given by $h=1$ with the origin node at 1-hop distance from the sink, and the upper bound $B_{Conv}^{max}$ by $h=4$ for $N=25$ ($h=6$ for $N=49$), with the origin node at a corner of the lattice network, and
\begin{equation}
B_{Conv}^{min}=B_{AODV}^{min}+B_{MAC}^{min},
\end{equation}
\begin{equation}
B_{Conv}^{max}=B_{AODV}^{max}+B_{MAC}^{max},
\end{equation}
for each source measurement.
For the proposed protocol, the number of bits obtained by simulations, may be expressed by $B_{Prop}=M P_{total,Prop}$, where $P_{total,Prop}$ is the total number of packets occuring in a session, including all the forwarded ones due to flooding.
Similarly, for the reference scheme with CDMA, we have $B_{CDMA}=NL P_{total,CDMA}$, as each sequence should have a minimal length of $NL$ to guarantee near-zero reconstruction error as explained in Section \ref{sec RelatedWork}.
Fig. \ref{fig:Bits} shows that the proposed protocol largely decreases the amount of packets per session in [kbytes] for networks with $25$ and $49$ nodes, corresponding to $80$ to $85\%$ reduction compared to the conventional algorithm with routing and MAC.
Compared to the reference scheme with CDMA, although the gain of our protocol is around $75\%$ for $N=25$, it becomes tremendously high for $N=49$, going up to $90\%$. Due to the large amount of bits required for having orthogonal sequences in the CDMA system, the total amount of packets drastically increases and even surpasses the conventional routing and MAC based protocol as the network enlarges, showing evident scalability issues that are alleviated by our proposed protocol.
Thus, the proposed protocol provides very low reconstruction errors, while achieving tremendous overhead savings that become even higher as the network enlarges.



\section{Conclusion}

We have proposed a novel protocol enabling non-scheduled, simultaneous transmissions for delivering measurement data to the sink in multi-hop wireless sensor networks.
Due to the sparse nature of the sensed events, the superimposed measurements resulting from interfered packet reception at each node can be recovered by CS algorithms. By making use of time stamps and signature sequences for each measurement, our protocol achieves low reconstruction errors, bringing significant savings in terms of overhead and energy, as compared to conventional routing and MAC scheduling strategies, as well as CDMA and flooding based protocol.

This fundamental work opens new vistas for further research. The protocol will be extended to more general systems, including path loss and channel fading, various network topologies, asynchronous nodes and, the more challenging case of analog transmissions. Another promising direction is the design of new algorithms based on the proposed ideas for general multi-hop mesh networks in the context of unicast/multicast transmissions.

\addcontentsline{toc}{chapter}{References}
\bibliographystyle{IEEEtran}
\bibliography{IEEEabrv,CS_Mesh_BIB}

\begin{thebibliography}{10}
\providecommand{\url}[1]{#1}
\csname url@rmstyle\endcsname
\providecommand{\newblock}{\relax}
\providecommand{\bibinfo}[2]{#2}
\providecommand\BIBentrySTDinterwordspacing{\spaceskip=0pt\relax}
\providecommand\BIBentryALTinterwordstretchfactor{4}
\providecommand\BIBentryALTinterwordspacing{\spaceskip=\fontdimen2\font plus
\BIBentryALTinterwordstretchfactor\fontdimen3\font minus
  \fontdimen4\font\relax}
\providecommand\BIBforeignlanguage[2]{{%
\expandafter\ifx\csname l@#1\endcsname\relax
\typeout{** WARNING: IEEEtran.bst: No hyphenation pattern has been}%
\typeout{** loaded for the language `#1'. Using the pattern for}%
\typeout{** the default language instead.}%
\else
\language=\csname l@#1\endcsname
\fi
#2}}

\bibitem{Can06feb}
{E.J. Candes, J. Romberg, T. Tao}, ``{Robust Uncertainty Principles: Exact
  Signal Reconstruction from Highly Incomplete Frequency Information},''
  \emph{{IEEE} Trans. on Info. Theory}, vol.~52, no.~2, pp. 489--509, February
  2006.

\bibitem{Don06apr}
{D.L. Donoho}, ``{Compressed Sensing},'' \emph{{IEEE} Trans. on Info. Theory},
  vol.~52, no.~4, pp. 1289--1306, April 2006.

\bibitem{Can06dec}
{E.J. Candes, T. Tao}, ``{Near-Optimal Signal Recovery From Random Projections:
  Universal Encoding Strategies?}'' \emph{{IEEE} Trans. on Info. Theory},
  vol.~52, no.~12, pp. 5406--5425, December 2006.

\bibitem{Hau08mar}
{J. Haupt, W.U. Bajwa, M. Rabbat, R. Nowak}, ``{Compressed Sensing for
  Networked Data},'' \emph{{IEEE} Signal Proc. Magazine}, vol.~25, no.~2, pp.
  92--101, March 2008.

\bibitem{Que09feb}
{G. Quer, R. Masiero, D. Munaretto, M. Rossi, J. Widmer and M. Zorzi}, ``{On
  the Interplay between Routing and Signal Representation for Compressive
  Sensing in Wireless Sensor Networks},'' in \emph{{ITA} {Workshop}}, San
  Diego, CA, February 2009.

\bibitem{Wan10oct}
{X. Wang, Z. Zhao, Y. Xia, H. Zhang}, ``{Compressed Sensing Based Random
  Routing for Multi-Hop Wireless Sensor Networks},'' in \emph{{IEEE} {ISCIT}},
  Tokyo, Japan, October 2010.

\bibitem{Ngu10oct}
{N. Nguyen, D.L. Jones and S. Krishnamurthy}, ``{Netcompress: Coupling Network
  Coding and Compressed Sensing for Efficient Data Communication in Wireless
  Sensor Networks},'' in \emph{{IEEE} {SIPS}}, Urbana, IL, October 2010.

\bibitem{Men09mar}
{J. Meng, H. li and Z. Han}, ``{Sparse Event Detection in Wireless Sensor
  Networks using Compressive Sensing},'' in \emph{{CISS}}, Baltimore, MD, March
  2009.

\bibitem{Mao10apr}
{R. Mao and H. Li}, ``{A Novel Multiple Access Scheme via Compressed Sensing
  with Random Data Traffic},'' in \emph{{IEEE} {WCNC}}, Sidney, Australia,
  April 2010.

\bibitem{Qas09sep}
{S.T. Qaseem, T.Y. Al-Naffouri, T.M. Al-Murad}, ``{Compressive Sensing Based
  Opportunistic Protocol for Exploiting Multiuser Diversity in Wireless
  Networks},'' in \emph{{IEEE} {PIMRC}}, Tokyo, Japan, September 2009.

\bibitem{Bha09jun}
{S.R. Bhaskaran, L. Davis, A. Grant, S. Hanly, P. Tune}, ``{Downlink Scheduling
  using Compressed Sensing},'' in \emph{{ITW}}, Volos, Greece, June 2009.

\bibitem{Elad10}
M.~Elad, \emph{Sparse and Redundant Representations}.\hskip 1em plus 0.5em
  minus 0.4em\relax Springer, 2010.

\bibitem{Dau04nov}
{I. Daubechies, M. Defrise and C. De-Mol}, ``{An Iterative Thresholding
  Algorithm for Linear Inverse Problems with a Sparsity Constraint},'' \emph{{}
  Comm. Pure Appl. Math.}, vol.~57, no.~11, pp. 1413--1457, November 2004.

\bibitem{aodv}
{C. Perkins, E. Belding-Royer and S. Das}, ``{Ad hoc On-Demand Distance Vector
  ({AODV}) Routing},'' http://www.ietf.org/rfc/rfc3561.txt, {IETF}, RFC 3561,
  July 2003.

\end{thebibliography}

\end{document}